\documentclass[]{iau} 
\usepackage{graphicx}
\usepackage{amsmath}

\title[Exceptional confined flares] 
{The Exceptional Aspects of the Confined X-class Flares
of Solar Active Region 2192}

\author[Julia K. Thalmann et al.] 
{Julia K. Thalmann$^1$, Yang Su$^{1,2}$, Manuela Temmer$^1$ \\
\and \\
Astrid M. Veronig$^1$}

\affiliation{
$^1$Institute of Physics/IGAM, University of Graz, \\ 
Universitätsplatz 5/II, 8010 Graz, Austria \\
email: {\tt julia.thalmann@uni-graz.at} \\[\affilskip]
$^2$Key Laboratory of Dark Matter \& Space Astronomy, \\
Purple Mountain Observatory, Chinese Academy of Sciences, \\
2 West Beijing Road, 210008 Nanjing, China \\
email: {\tt yang.su@pmo.ac.cn}
}

\pubyear{2015}
\volume{S320} 
\jname{Solar and Stellar Flares and Their Effects on Planets}
\editors{A.G. Kosovichev, S.L. Hawley \& P. Heinzel, eds.}

\begin{document}

\maketitle

\begin{abstract}
During late October 2014, active region NOAA~2192 caused an unusual high level of 
solar activity, within an otherwise weak solar cycle. While crossing the solar disk, 
during a period of 11 days, it was the source of 114 flares of {\it GOES} class C1.0 
and larger, including 29 M- and 6 X-flares.
Surprisingly, none of the major flares ({\it GOES} class M5.0 and larger) was 
accompanied by a coronal mass ejection, contrary to statistical tendencies found in
the past. From modeling the coronal magnetic field of NOAA~2192 and its surrounding, 
we suspect that the cause of the confined character of the
flares is the strong surrounding and overlying large-scale magnetic field. 
Furthermore, we find evidence for multiple magnetic 
reconnection processes within a single flare, during which electrons were 
accelerated to unusual high energies.

\end{abstract}

\begin{figure}
	\begin{center}
		\includegraphics[width=0.75\columnwidth]{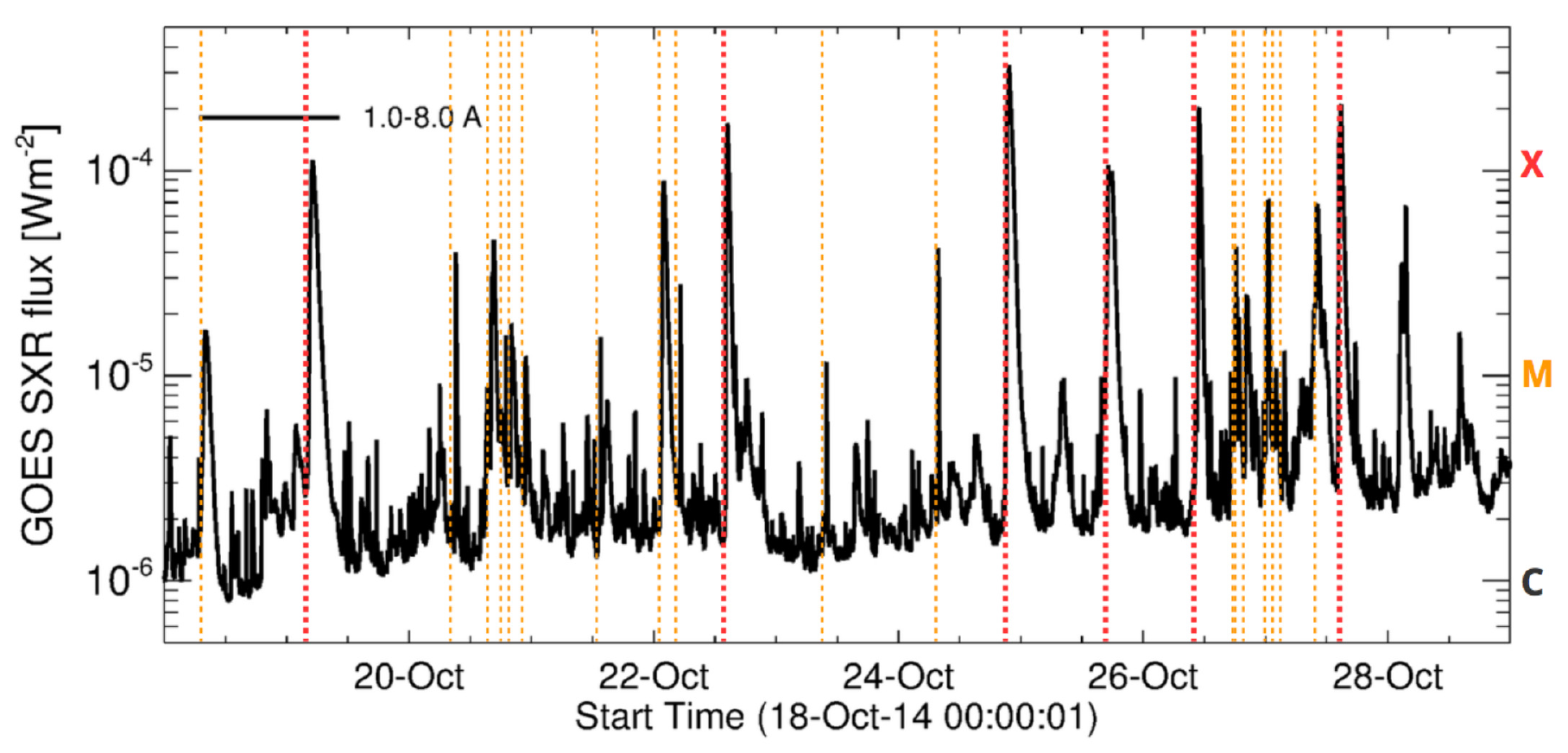} 
		\caption{{\it GOES} 1.0--8.0~\AA\ soft X-ray flux between October 18 and 29, 
		2014. Vertical dashed lines indicate the time when a flare occurred. Yellow and 
		red lines mark the peak time of M- and X-flares, respectively. Only peak times of
		flares M1.0 and larger are shown. All indicated flares originated from NOAA~2192.}
		\label{fig1}
	\end{center}
\end{figure}

\begin{figure}
	\begin{center}
		\includegraphics[width=0.9\columnwidth]{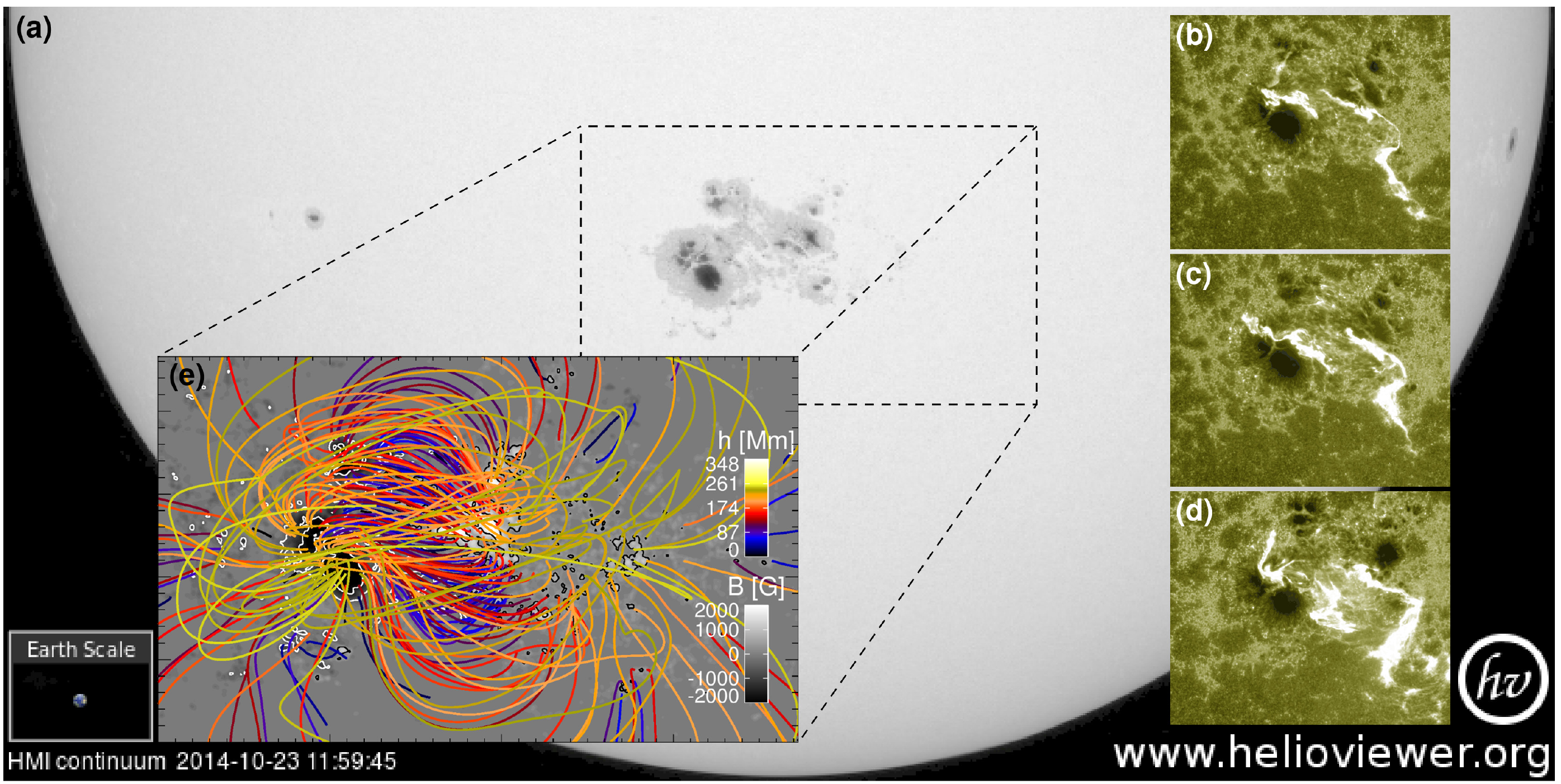} 
		\caption{(a) {\it SDO}/HMI white-light continuum image on October 23, 
		2014 at 12:00~UT, covering the southern hemisphere 
		of the Sun. For an impression of the size of NOAA~2192, 
		Earth's size is indicated in the lower left corner. Panels (b)--(d) show 
		the {\it SDO}/AIA 1600~\AA\ emission in the active region center at the 
		peak time of three X-flares, showing a clearly similar morphology.
		Panel (e): Sample field lines calculated from a nonlinear force-free magnetic field 
		model, colored according to the apex height of each field line. The active region 
		center consists of a system of highly sheared fields (black/blue/purple lines).}
		\label{fig2}
	\end{center}
\end{figure}

We analyze the unexpected high solar activity associated to active region NOAA~2192 
which hosted more than one hundred flares during disk passage, including 6
{\it GOES} class X flares (Fig.\,\ref{fig1}). NOAA~2192 covered a large part of the 
solar surface (roughly 15--20 times Earth's diameter in east-west direction; see 
Fig.\,\ref{fig2}a) and the intensive flaring activity was rooted in its complex 
and strong surface magnetic field configuration. It was classified as 
$\beta\gamma\delta$-configuration and hosted umbral field strengths of $>2.5$~kG 
(the latter corresponding to the upper end of known statistics of umbral magnetic field 
strengths). 

The flaring activity of NOAA~2192 was exceptional in that all of the large ({\it GOES} 
class $\ge$M5.0) flares were confined, i.\,e., that no associated ejection of coronal 
material into interplanetary space was observed. This is special since we know of the 
affinity of large flares and coronal mass ejections to occur together (see 
\cite[Yashiro et al. 2006]{Yashiro_etal06}). 
In \cite[Thalmann et al. (2015)]{Thalmann_etal15}, we analyzed some 
exceptional aspects of the major flares that originated from 
NOAA~20192 on October 22 and 24, 2014. 
In the following, we summarize the most important findings.

{~\\ \centering \it 1.~Magnetic reconnection at large heights in the corona\\~\\}

All of the analyzed major flares were obviously similar in morphology 
(see Fig.\,\ref{fig2}b--d). The observed flare ribbons showed a large initial 
separation and no substantial growth of separation at later times 
(cf.~Fig.\,1 of \cite[Thalmann et al. 2015]{Thalmann_etal15}). That suggests that the 
reconnection site was situated at large coronal heights ($\approx$50~Mm above 
photospheric levels) during these events. That corresponds
roughly to the average pre-flare apex height of the strongly sheared magnetic field 
structures in the center of the active region,
calculated from a nonlinear force-free magnetic field model (see Fig.\,\ref{fig2}e).

{~\\ \centering \it 2.~Multiple magnetic reconnection events during a single flare\\~\\}

\begin{figure}
	\begin{center}
		\includegraphics[width=0.9\columnwidth]{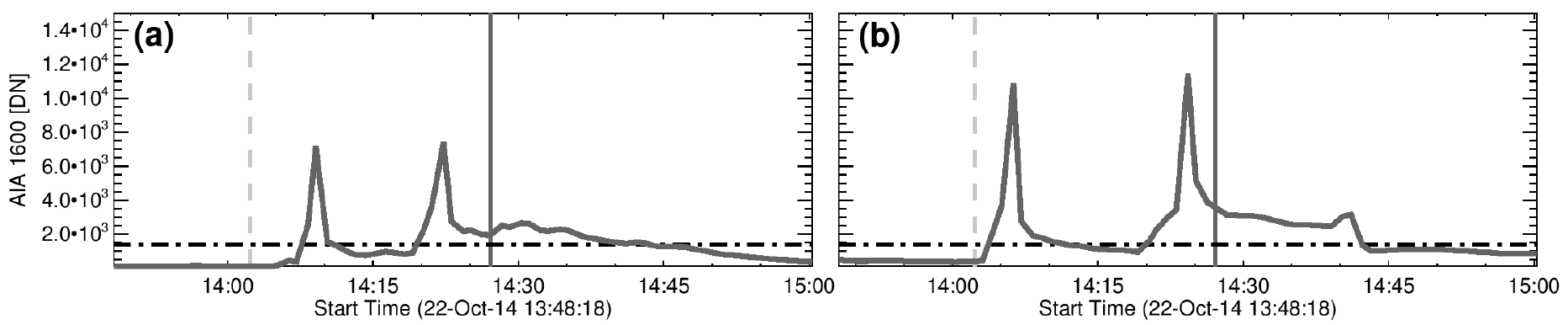} 
		\caption{{\it SDO}/AIA 1600~\AA\ light curves, typical for localized kernels within 
		flare ribbons located in the (a) negative and (b) positive polarity domain of
		NOAA~2192. The vertical dashed/solid lines mark the nominal start and 
		peak time of the X1.6 flare, respectively, based on the {\it GOES} soft X-ray flux. 
		The horizontal dotted line indicates the intensity threshold used to track flaring pixels.}
		\label{fig3}
	\end{center}
\end{figure}

Analysis of the flare ribbon emission during the X1.6 flare, peaking on October 
22 at 14:28~UT, revealed that the extreme ultraviolet light curves of a number of flare 
kernels (i.\,e., the constituents of the typical ribbon-like flare emission) exhibited two 
distinct peaks. Examples of typical light curves of flare kernels located in the negative 
and positive polarity region of NOAA~2192 are shown in Fig.\,\ref{fig3}
(see also Fig.\,5 of \cite[Thalmann et al. 2015]{Thalmann_etal15}).
Importantly, those peaks occurred co-temporal with two hard X-ray bursts 
(compare.~Fig.\,3 of \cite[Thalmann et al. 2015]{Thalmann_etal15})
and co-spatial with H$\alpha$ flare kernels and non-thermal X-ray sources
(cf.~Fig.\,4 of \cite[Thalmann et al. 2015]{Thalmann_etal15}). 
These findings provide evidence that multiple reconnection events can occur within a 
single flare, i.\,e., that the same magnetic field structures can be involved in successive
reconnection events. 
The {\it RHESSI} hard X-ray data revealed steep non-thermal power-law spectra of the 
flare-accelerated electrons (cf.~Fig.\,3b--3e of \cite[Thalmann et al. 2015]{Thalmann_etal15}). 
The total energy in electron derived from the spectra 
indicated an unusually large amount of total kinetic energy that resided in the 
accelerated electrons, in contrast to typical (eruptive) flares of class X1. 

{~\\ \centering \it 3.~External magnetic field constraining the development of mass ejections\\~\\}

\begin{figure}
	\begin{center}
		\includegraphics[width=0.75\columnwidth]{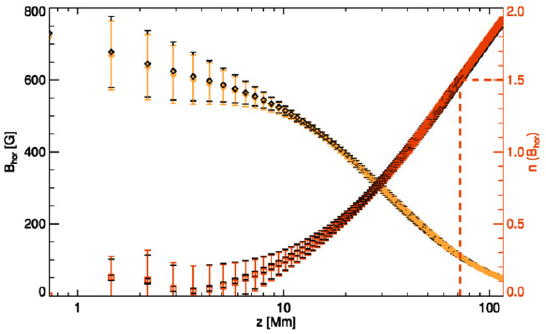} 
		\caption{Average magnitude of the horizontal magnetic field above the main 
		polarity inversion line of NOAA~2192 prior to and after the X3.1 flare on October 24. 
		Diamonds and squares mark the average horizontal field magnitude,
		$\mathit{B_{\rm hor}}$, and the decay index $n$, respectively. 
		Black symbols refer to the pre-flare and colored symbols to the
		post-flare magnetic field configuration.}
		\label{fig4}
	\end{center}
\end{figure}

Modeling of the large-scale (potential) coronal neighborhood of NOAA~2192, around 
the X3.1 flare (peak time 21:41~UT on October 24) revealed a north-south oriented magnetic
arcade on top of the flaring region (cf.~Fig.\,2 of \cite[Thalmann et al. 2015]{Thalmann_etal15}).
We estimated the average strength of the overlying field and its decay index $n$ 
with height (i.\,e., a measure for the decay of the constraining background field; see 
Fig.\,\ref{fig4}). We stress that, by definition, these values 
have to be approximated from a potential (current-free) magnetic field configuration.
For values of $n\gtrsim1.5$, theory predicts favorable conditions for the onset of torus
instability. Hence, the external field has to decrease sufficiently fast in the direction 
of the major radius of a torus carrying a toroidal current (\cite[Kliem \& T{\"o}r{\"o}k 2006]{Kliem_Toeroek_06}).
In the present case, this was found only true for heights of $\sim$70~Mm above
the solar surface, i.\,e., at much larger heights than the apexes of the highly sheared 
core-fields, involved in the flaring process. Also \cite[Sun et al. (2015)]{Sun_etal15}
studied the evolution of NOAA~2191 and pointed out the strength of the overlying 
field. In addition they found that this CME-less active region revealed a weaker 
non-potentiality and smaller flare-related magnetic field changes than 
CME-associated flaring regions.

Note that the concept of torus instability is well defined only for a configuration 
consisting of a well-defined, narrow current following the axis of a flux tube (the current-carrying torus). 
The applicability of this concept, to complex magnetic
field configurations such as that of NOAA~2192 is not straight forward, however,
since it consists of widely spread electric currents. Thus, we also employed a 
more general measure for the strength of the constraining
background field, the flux ratio, in the form $F_{\rm low}/F_{\rm high}$ above the 
polarity inversion line. Here, $F_{\rm low}$ is the average 
horizontal flux in the height range $1.0<h<1.1~R_{\rm Sun}$ (covering the 
sheared core field) and $F_{\rm high}$ is the corresponding flux within 
$1.1<h<1.5~R_{\rm Sun}$ (covering the overlying arcade that may prevent the
underlying core field from erupting).
We found$F_{\rm low}/F_{\rm high}\approx 0.3$, a value even lower than the flux 
ratios found by \cite[Wang \& Zhang (2007)]{Wang_Zhang_07} for confined X-flares
($1.0\lesssim F_{\rm low}/F_{\rm high}\lesssim 6.0$).

To summarize, we find that the strength of the coronal fields in the neighborhood of 
active regions may be essential in determining whether or not the upcoming flaring 
activity is associated to a mass ejection or not. That means that a strong background
field may prevent an otherwise unstable flux rope from erupting. 
Our nonlinear force-free magnetic field modeling, however, did not reveal the presence of a fully emerged,
well-defined flux rope, that could be subject to, e.\,g., torus instability after all. Instead,
it shows a system of rather strongly sheared magnetic fields 
(see Fig.\,\ref{fig2}e). This is supported by the observation of 
\cite[Veronig \& Polanec (2015)]{Veronig_Polanec_15}, who did not find evidence for 
an existing filament that would indicate the presence of a flux rope in the course 
of the X1.6 flare on October 22. Therefore, they 
put forward an emerging-flux scenario as an explanation for the associated large but 
confined flaring activity.

\acknowledgments The authors thankfully acknowledge support from the Austrian 
Science Fund (FWF): P25383-N27, P27292-N20, and V195-N16.

\end{document}